\documentclass[12pt]{article} 
\usepackage{cite}
\usepackage{graphicx}
\usepackage{subfig}

\usepackage{color}
\usepackage{psfrag}

\begin{document}
\def \d {{\rm d}}
\def\op#1{\mathord{\rm #1}}

\title{Chaotic motion in Kundt spacetimes}

\author{
J. Podolsk\'y\thanks{E--mail: {\tt podolsky`AT'mbox.troja.mff.cuni.cz}}
,\  D. Kofro\v n\thanks{E--mail: {\tt kofrd0am`AT'artax.karlin.mff.cuni.cz}}
\\ \\ Institute of Theoretical Physics,\\
Faculty of Mathematics and Physics, Charles University in Prague,\\
V Hole\v{s}ovi\v{c}k\'ach 2, 180~00 Prague 8, Czech Republic.\\ }

\maketitle
\baselineskip=19pt

\begin{abstract}
We demonstrate that geodesics in exact vacuum Kundt gravitational waves may exhibit a highly complicated behaviour. In fact, as in the previously studied case of non-homogeneous {\it pp}-waves, for specific choices of the structural function the motion appears to be genuinely chaotic. This fact is demonstrated by the fractal method.
\end{abstract}

\vfil\noindent

\vfil
\eject

\section{Introduction}

Chaotic motions were already found and described in a number of various spacetimes, in particular in the  black hole systems, see e.g.  \cite{Conto1,BoCa,DFC,Yur,CG,LeVi,SuMa,Le1,Le2,CL,Hartl1,Hartl2,HaBu}, or in special axisymmetric solutions \cite{KaVo,SSM,ViLe,ViLe2}. Some time ago we discovered chaotic behaviour of geodesics in specific vacuum {\it pp}-waves \cite{PVcha,PoVe2,PoVe3,VePo}. Although chaotic interaction of test particles with linearized gravitational waves propagating on flat Minkowski background had been studied before \cite{BoCa,LeVi,VaPa,CF,CMR,Koku}, this was the first explicit demonstration of chaos in \emph{exact} radiative spacetimes.

A natural question thus emerged whether a similar complicated behaviour of geodesics could be found in other exact solutions of Einstein's equations that represent gravitational waves. There exist several important classes of such solutions which are geometrically different from  {\it pp}-waves, for example the family of Kundt spacetimes with non-expanding wavefronts or radiative spacetimes of the Robinson--Trautman type \cite{KSMH} which describe expanding waves.

In the present work we concentrate on specific vacuum type~N spacetimes from the class of solutions that admit a non-expanding, shear-free, and twist-free null geodesic congruence which was first described in 1961 by Wolfgang Kundt  \cite{Kundt61,Kundt62,EK} (for recent reviews see, e.g., \cite{KSMH,PodBel04}). These so-called Kundt waves may be interesting from the theoretical point of view because all curvature invariants of all orders identically vanish for such spacetimes \cite{BicakPravda,Pravdovi}. They may thus play an important role in string theory and quantum gravity since there are no quantum corrections to all perturbative orders \cite{Coley}.

Some aspects of geodesic motions in these spacetimes have already been studied and described. For example, it was shown in \cite{BicPod99II} that geodesic deviations exhibit a transverse structure of relative motions of test particles. In \cite{PodBel04} we presented explicit families of timelike, null and spacelike geodesics and parallelly propagated frames which enabled us to elucidate some global properties of the spacetimes, such as an inherent rotation of the wave-propagation direction, or the character of singularities. In particular, we demonstrated that the characteristic envelope singularity of the rotated wave-fronts \cite{GriPod98} is a (non-scalar) curvature singularity, although all scalar invariants of the Riemann tensor vanish there.

Nevertheless, an open question has remained whether geodesic motion itself exhibits chaotic behaviour, analogous to the chaos in {\it pp}-waves \cite{PVcha,PoVe2,PoVe3,VePo}. In the present work we demonstrate that this is indeed the case. After a brief review of the Kundt vacuum waves in section~\ref{sec2}, we show in section~\ref{sec3} by a numerical fractal method that geodesics in these spacetimes are  chaotic.

\section{Kundt vacuum spacetimes of type N}
\label{sec2}
The metric of the Kundt gravitational waves \cite{Kundt61,Kundt62,EK,KSMH} (distinct from the widely known class of {\it pp}-waves)
can be written in the form \cite{PodBel04}
\begin{equation}
{\d s}^2={\d x}^2+{\d y}^2-4x^2\d u\,\d v +4(x^2v^2+xG)\,{\d u}^2\ ,
\label{metric}
\end{equation}
where the function $G(x,y,u)$ depends on retarded time $u$ and on spatial coordinates ${x,y}$ which span the plane wave-fronts ${u=\hbox{const}}$. Due to the presence of a (non-scalar) curvature ``envelope'' singularity at ${x=0}$, we have to restrict ourselves to the region in which ${x>0}$. The geodesic equations corresponding to (\ref{metric}) are
\begin{eqnarray}
&&\ddot{x}=-4x\dot{v}\dot{u}+(4xv^2+2G+2xG_{,x})\dot{u}^2\ ,\label{gx}\\
&&\ddot{y}=2xG_{,y}\dot{u}^2\ ,\label{gy}\\
&&\ddot{u}=-2\frac{\dot{x}}{x}\dot{u}-2v\dot{u}^2\ ,\label{gu}\\
&&\ddot{v}=-2\frac{\dot{x}}{x}\dot{v}+4v\dot{v}\dot{u}+2\left(\frac{G}{x}\right)_{\!,x}\!\dot{x}\dot{u}
+2\left(\frac{G}{x}\right)_{\!,y}\!\dot{y}\dot{u}+\left[\left(\frac{G}{x}\right)_{\!,u} \!\!-4v^3-4v\frac{G}{x}\right]\dot{u}^2\ ,\label{gv}
\end{eqnarray}
where the dot denotes differentiation with respect to an affine parameter $\tau$. In addition, there is the normalization condition of
four-velocity:
\begin{equation}
\dot{x}^2+\dot{y}^2-4x^2\dot{u}\dot{v}+4(x^2v^2+xG)\,\dot{u}^2=\epsilon\ ,
\label{Normalization}
\end{equation}
where ${\epsilon=-1,0,+1}$ denotes respectively the timelike, null or spacelike character of the geodesic.

Specific geodesic motion given by  equations (\ref{gx})--(\ref{Normalization}) crucially depends on the particular form of the spacetime function $G(x,y,u)$. We concentrate here on type~N vacuum Kundt spacetimes~(\ref{metric}). It such a case the function ${G\not=0}$ satisfies the Laplace equation ${G_{,xx}+G_{,yy}=0}$, see \cite{PodBel04}. A general solution to this field equation is thus ${G=\op{Re}\{g(\xi, u)\}}$, where  $g$ is an arbitrary function holomorphic in the argument ${\xi=x+\hbox{i}\, y}$. If the function $g$ is linear in $\xi$ or independent of it, this generates only a flat space. The simplest radiative Kundt spacetimes thus arise when ${g=d(u)(\xi-\xi_s)^n}$, ${n=2,3,4,\ldots\,}$, ${\xi_s=\hbox{const}}$. As in \cite{PVcha,PoVe2} we will consider only the waves of a constant profile, i.e. $G$ independent of $u$, so that ${d(u)=\lambda\equiv\hbox{const.}}$ We can then use the translational symmetry in $y$ to remove the constant $y_s$. We will thus investigate geodesics in ``polynomial'' Kundt spacetimes characterized by the structural function
\begin{equation}
G=\lambda\,\op{Re}\{(x-x_s+\hbox{i}\, y)^n\}\ ,
\label{GObecna}
\end{equation}
where $\lambda,x_s$ are constants, and ${n=2,3,4,\cdots\,}$. Introducing  polar coordinates,
${x-x_s=r\cos\varphi}$, ${y=r\sin\varphi}$, we may write ${G=\lambda\,r^n\cos(n\varphi)}$
which is an $n$-saddle function.

\begin{figure}[ht]
\centering 
\subfloat[][$n=6$]{\includegraphics[width=4cm,keepaspectratio]{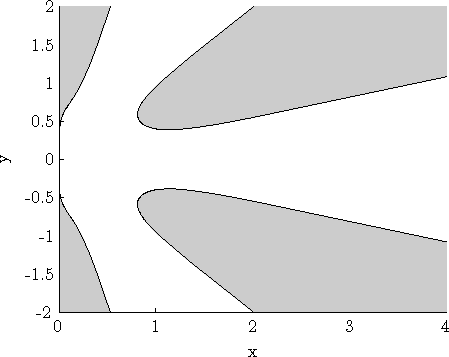}}\qquad
\subfloat[][$n=7$]{\includegraphics[width=4cm,keepaspectratio]{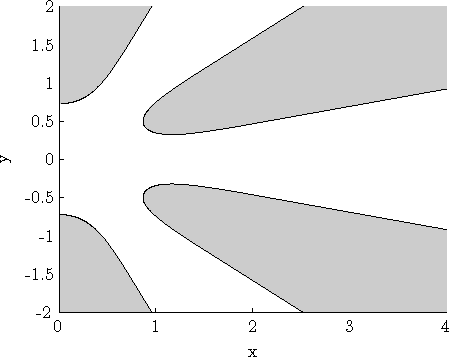}}\qquad
\subfloat[][$n=8$]{\includegraphics[width=4cm,keepaspectratio]{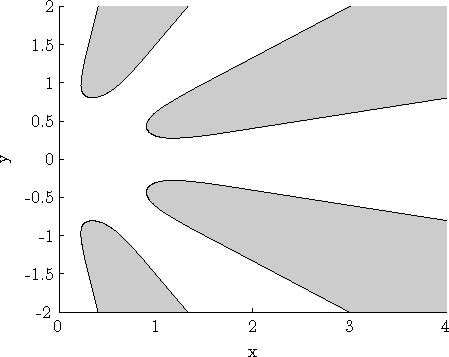}}

\subfloat[][$n=6$]{\includegraphics[width=4cm,keepaspectratio]{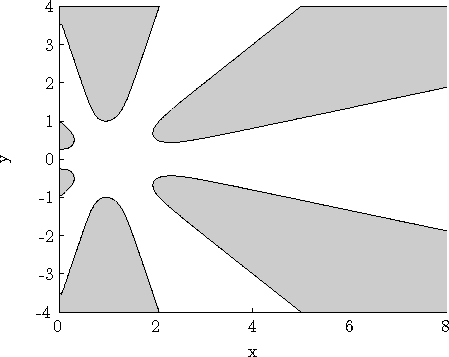}}\qquad
\subfloat[][$n=7$]{\includegraphics[width=4cm,keepaspectratio]{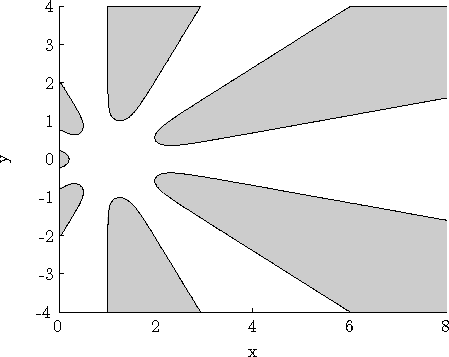}}\qquad
\subfloat[][$n=8$]{\includegraphics[width=4cm,keepaspectratio]{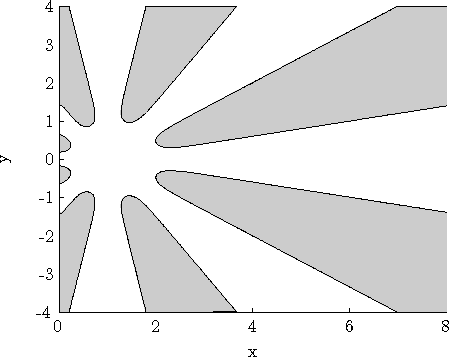}}
%  *.png -> *.eps
\caption{Regions of different character of the Killing
vector $\partial_u$: in the gray regions it is timelike, while in
the white regions it is spacelike. On  the boundaries the Killing
vector is null. The upper three pictures correspond to ${x_s=0}$, the lower ones correspond to ${x_s=1}$.} \label{f:kill-re}
\end{figure}

The metric (\ref{metric}), (\ref{GObecna}) admits the Killing vector $\partial_u$ whose norm is given by 
${g_{uu}=4(x^2v^2+xG)}$. These Kundt spacetimes can thus be invariantly divided into regions in which this Killing vector is timelike or spacelike, separated by the boundaries on which the vector $\partial_u$ in null. This appears when $xv^2$ is smaller tan, greater than, or equal to $-G(x,y)$, respectively. For a fixed value of~$v$ this is visualized in figure~\ref{f:kill-re} for ${n=6,7,8}$: the upper pictures correspond to ${x_s=0}$ while the lower ones correspond to ${x_s=1}$. The Killing vector is timelike in the gray regions while it is spacelike in the white regions of the $(x,y)$-plane.

\section{Demonstration of chaos in polynomial Kundt waves}
\label{sec3}

The system of non-linear differential equations (\ref{gx})--(\ref{gv}) with (\ref{GObecna}) is quite complicated, so one can explicitly find only a few very special analytic solutions, see \cite{PodBel04}. In order to obtain the global picture of a geodesic motion in the studied spacetimes, it is necessary to perform numerical investigations. 

\subsection{The choice of initial data}
\label{ssec1}

In these numerical integrations we make the following choice of initial conditions. In the Kundt coordinates ${(x,y,u,v)}$, the initial \emph{position} of each geodesic $x^\mu(\tau)$ is given by
\begin{equation}
x^\mu(0)=(x_0, y_0, 0, x_0^{-1})\ ,
\label{inipos}
\end{equation}
where
\begin{equation}
\textstyle{x_0=x_s+r_0\cos\varphi\ ,\qquad y_0=r_0\sin\varphi}\ ,
\label{inipos2}
\end{equation}
while the initial \emph{velocity} is
\begin{equation}
\frac{\d x^\mu}{\d\tau}(0)=\left(-\frac{1}{x_0}, 0, \frac{1}{2x_0}, \frac{1}{2x_0}+\frac{1}{x_0^3}+\frac{G(x_0,y_0)}{2x_0^2}\right)\ .
\label{inivel}
\end{equation}
For given $x_s$ and $r_0$, it is thus a one-parameter family of geodesics parametrized by $\varphi$.

\begin{figure}[ht]
\centering 
\subfloat[][${G=\lambda\,\op{Re}\{(x+\hbox{i}\, y)^8\}}$]{\includegraphics[width=7cm,keepaspectratio]{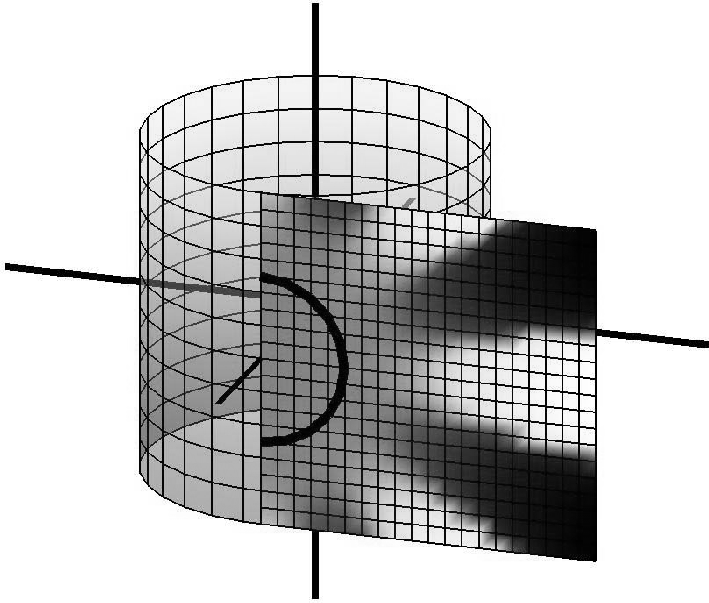}}\qquad 
\subfloat[][${G=\lambda\,\op{Re}\{(x-1+\hbox{i}\, y)^8\}}$]{\includegraphics[width=7cm,keepaspectratio]{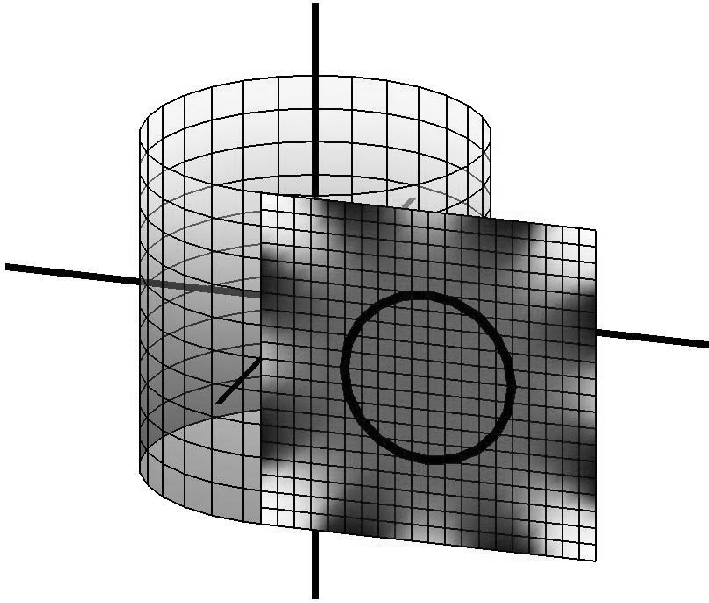}} 
%  *.png -> *.eps
\caption{Initial positions of test particles are localized on a circle of radius $r_0$ in the planar wave-surface ${u=0}$ which is tangent to an expanding cylinder with the vertical axis~$Y$ and transverse directions $X,Z$. The particles are at rest in Minkowski space in front of the Kundt (shock) wave. The value of the metric function $G(x,y)$ on the initial wave-surface is also indicated by various levels of grey (cf figure~\ref{f:kill-re}).}
\label{f:poly-ini}
\end{figure}

The geometric meaning and justification of such ``somewhat peculiar'' choice of the initial conditions is provided by the transformation from the Kundt coordinates ${(x,y,u,v)}$ to the standard Minkowski coordinates ${(X,Y,Z,T)}$ for the flat space case when  ${G=0}$, namely
\begin{eqnarray}
&&X=x(1+2uv)\ ,\hspace{17mm} Y=y\ ,\nonumber\\
&&Z=x[v-u(1+uv)]\ ,\qquad T=x[v+u(1+uv)]\ ,\label{transfMink}
\end{eqnarray}
see \cite{GriPod98,PodBel04}. We immediately obtain
\begin{equation}
X^\mu(0)=(x_0, y_0, 1, 1)\ ,\qquad
\frac{\d X^\mu}{\d\tau}(0)=(0, 0, 0, 1)\ .
\label{iniposMink}
\end{equation}
All these test particles thus \emph{start from rest} (at the same time ${T=1}$ and plane ${Z=1}$) from the position $(X,Y)=(x_0,y_0)$, i.e. on a \emph{circle} of radius $r_0$, centred around ${X=x_s}$, ${Y=0}$. The parameter $\varphi$ in (\ref{inipos2}) is the standard polar angle along this circle.  In view of the geometric interpretation of the Kundt spacetimes presented in \cite{GriPod98,PodBel04}, the initial positions of the geodesic test particles are localized on ${u=0}$, which is a half-plane tangent to an expanding cylinder ${X^2+Z^2-T^2\equiv x^2=0}$, $Y$ arbitrary, as illustrated in figure~\ref{f:poly-ini}.

The last term in expression (\ref{inivel}), which is proportional to $G$, has been included in order to guarantee the consistency of the geodesic equations and the normalization condition (\ref{Normalization}) across the wave-surface ${u=0}$ even in a non-flat case. In other words, we assume the continuity of the geodesics in the Kundt coordinates ${(x,y,u,v)}$ on ${u=0}$, and also the continuity of the corresponding coordinate velocities, except for an inevitable jump in $\dot{v}$ equal to ${(G/x)}\dot{u}$. The~physical interpretation of these assumptions is that, in fact, we investigate geodesic motion of free test particles which are initially at rest in a (locally) Minkowski space in front of a \emph{shock wave}. These particles are hit simultaneously by the initial wave-front ${u=0}$, and subsequently they start to move under the influence of the Kundt gravitational wave described by (\ref{metric}), (\ref{GObecna}).

\subsection{The geodesics}
\label{ssec2}

Let us now present the numerically obtained results which demonstrate a typical behaviour of geodesics in the above ``polynomial'' Kundt spacetimes.

\begin{figure}[ht]
\centering 
\subfloat[][]{\includegraphics[width=6cm,keepaspectratio]{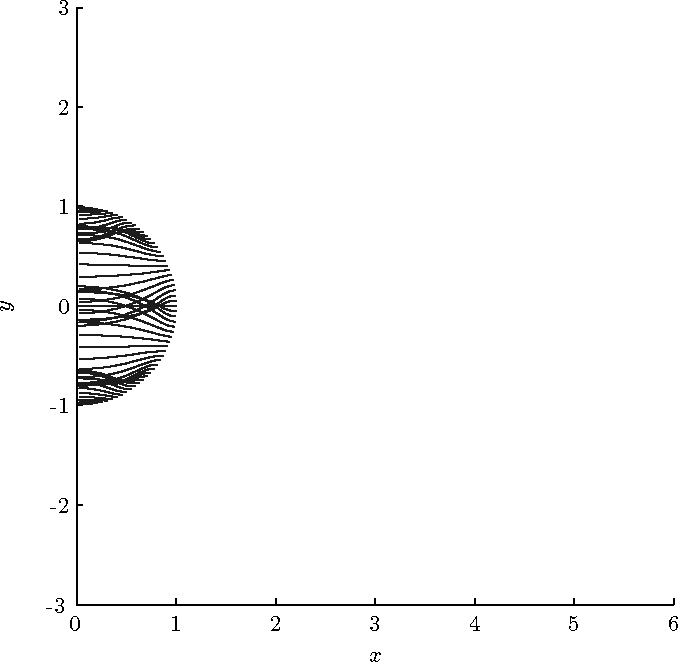}}\quad
\subfloat[][]{\includegraphics[width=6cm,keepaspectratio]{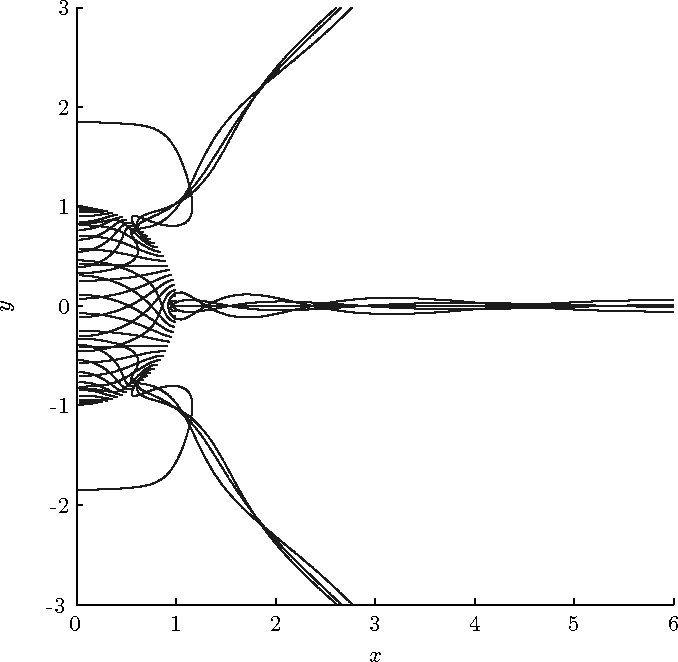}}
\vskip6mm
\subfloat[][]{\includegraphics[width=6cm,keepaspectratio]{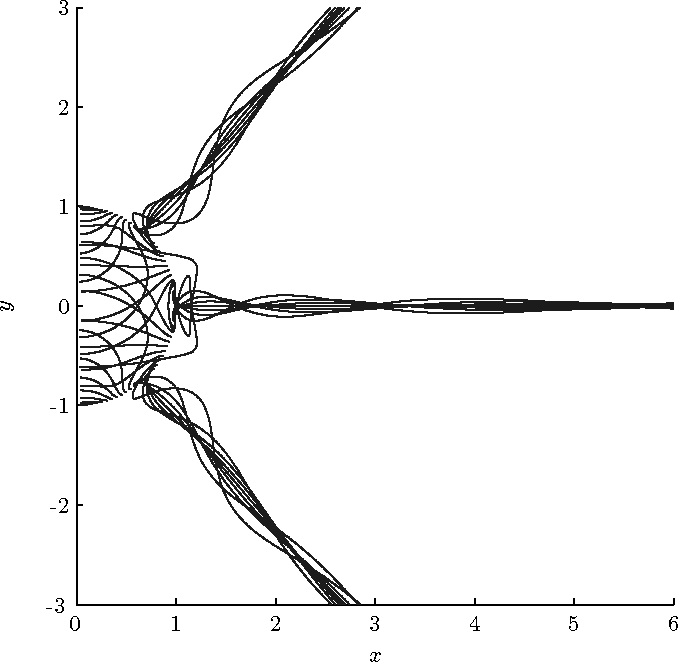}}\quad
\subfloat[][]{\includegraphics[width=6cm,keepaspectratio]{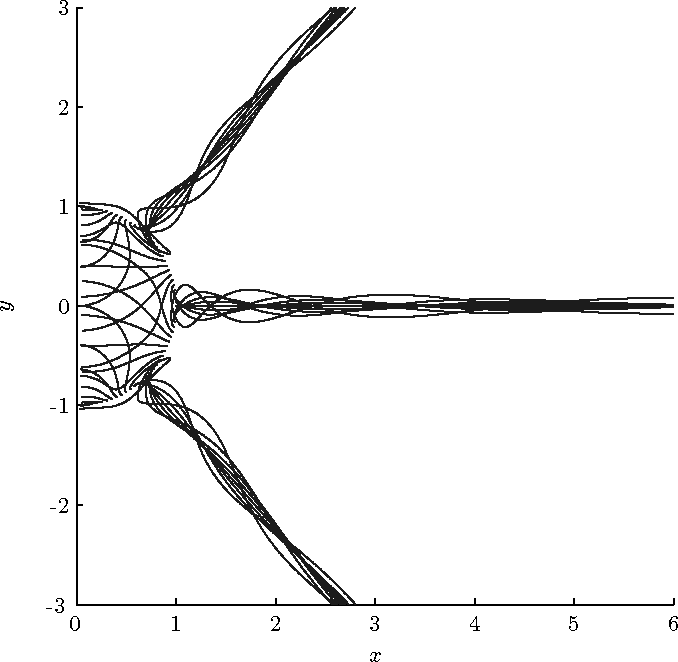}}
\caption{Trajectories of the test particles for ${G=\lambda\,\op{Re}\{(x+\hbox{i}\, y)^8\}}$.} \label{f:r8-0}
\end{figure}

In figure~\ref{f:r8-0}, we show a family of geodesics in the spacetimes (\ref{metric}), (\ref{GObecna}) with ${x_s=0}$, ${n=8}$, i.e.
${G=\lambda\,\op{Re}\{(x+\hbox{i}\, y)^8\}}$,  for initial data (\ref{inipos})--(\ref{inivel}), ignoring here the value of the coordinates $u$ and $v$. We have chosen the characteristic  radius ${r_0=1}$ and considered the cases when ${\lambda=2,4,16,32}$ which are shown in parts (a), (b), (c), (d).

%  *.png -> *.eps

We observe that the geodesics either (in a finite proper time) terminate in the envelope singularity ${x=0}$, or are unbounded and \emph{radially escape to infinity} in the $(x,y)$-plane. Interestingly, they escape {\it through only one of the outcome channels} which exactly correspond to the axes of the regions in which the Killing vector $\partial_u$ is spacelike, cf the white regions in figures~\ref{f:kill-re} and~\ref{f:poly-ini}. As in the case of non-homogeneous {\it pp}-waves \cite{PVcha,PoVe2,PoVe3,VePo}, as the test particles escape, their frequency of oscillations around the axis of the outcome channel grows to infinity while the amplitude of  oscillations tends to zero. For small values of the parameter $\lambda$, the ``attractive effect'' of the envelope singularity is dominant so that almost all particles approach ${x=0}$ while, with a growing value of $\lambda$, the effect of the gravitational wave becomes stronger and thus more and more geodesic particles are forced to escape to infinity through the outcome channels. In fact, they are attracted to the curvature singularity at ${r=\infty}$, see expression (\ref{GObecna}) which has the form ${G=\lambda\,r^n\cos(n\varphi)}$.

\begin{figure}[ht]
\centering 
\subfloat[][]{\includegraphics[width=6cm,keepaspectratio]{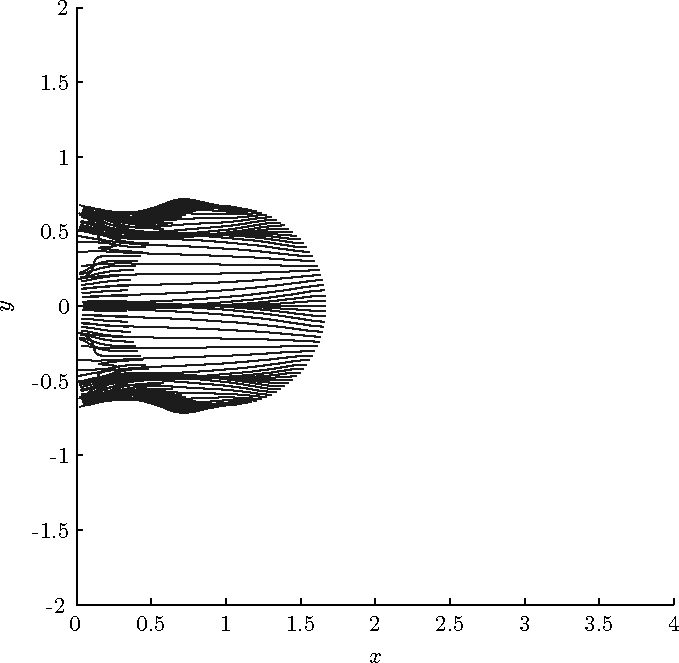}}\quad
\subfloat[][]{\includegraphics[width=6cm,keepaspectratio]{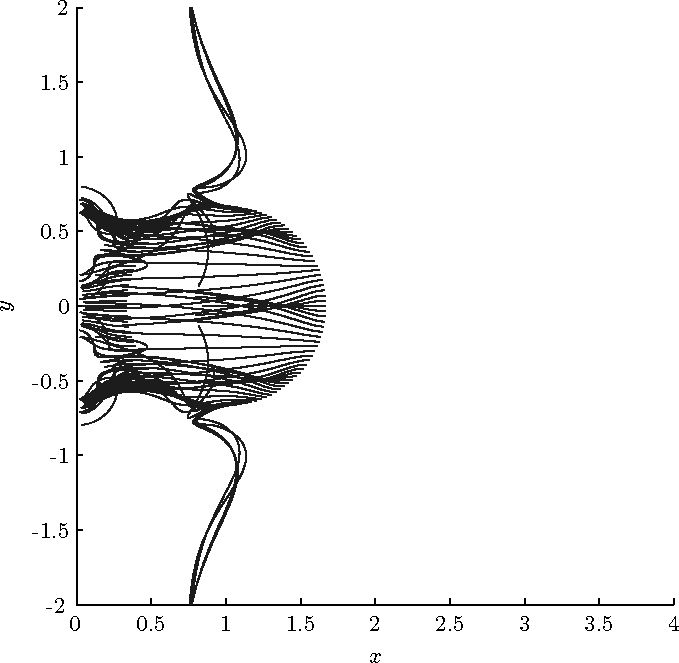}}
\vskip6mm
\subfloat[][]{\includegraphics[width=6cm,keepaspectratio]{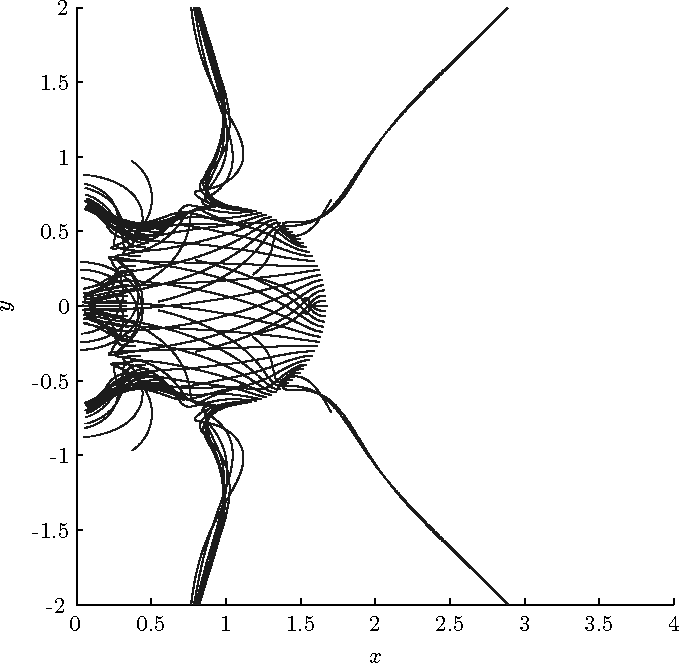}}\quad
\subfloat[][]{\includegraphics[width=6cm,keepaspectratio]{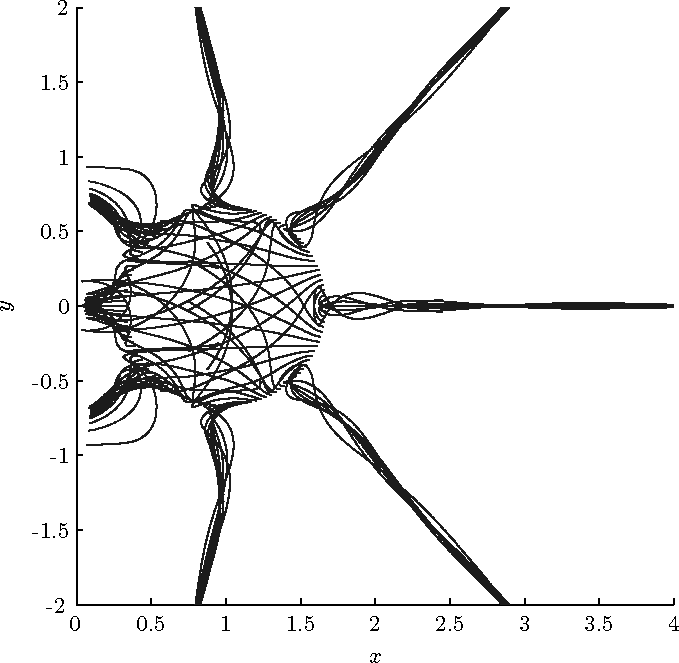}}
\caption{Trajectories of the test particles for ${G=\lambda\,\op{Re}\{(x-1+\hbox{i}\, y)^8\}}$.} \label{f:r8-1}
\end{figure}

This characteristic behaviour also occurs in the case ${G=\lambda\,\op{Re}\{(x-1+\hbox{i}\, y)^8\}}$ corresponding to  ${x_s=1}$, ${n=8}$, with the radius of the initial circle ${r_0=\frac{2}{3}}$,  illustrated for ${\lambda=8,16,32,64}$ in parts (a), (b), (c), (d) of figure~\ref{f:r8-1}, respectively. Again, for small values of the parameter $\lambda$ all the geodesics end in the envelope singularity ${x=0}$, but with a growing $\lambda$ the effect of the polynomial Kundt gravitational wave increases and the outcome channels open. These coincide with the regions in which the Killing vector $\partial_u$ is spacelike, see figures~\ref{f:kill-re},~\ref{f:poly-ini}. Although  the geodesics which move through the channels with decreasing $x$ still terminate at ${x=0}$ (in fact, such channels are somewhat ``bent'' by the presence of the envelope singularity), those with a growing $x$ escape to infinity, $r\to\infty$.

\subsection{The fractal structure of geodesic motion}
\label{ssec3}

The main objective of the present paper, however, is to establish the
chaotic behaviour of geodesics in the Kundt-wave spacetimes.

Chaos is usually indicated by a highly sensitive dependence of possible 
outcomes on the  choice of initial conditions. The standard invariant approach to prove chaotic motion,
called a fractal method, was advanced in general relativity, e.g.,  
in the papers \cite{Conto1,DFC,Yur,CG}. It starts with a definition of several
different outcomes, that is ``types of ends'' of all possible
trajectories. Subsequently, a large set of initial conditions is
evolved numerically until one of the outcome states is always
reached. Chaos is established if the basin boundaries which separate
initial conditions leading to different outcomes are fractal.
We shall now demonstrate that exactly these structures can be observed
in the system studied.

\begin{figure}[ht]     % -- level 1 -----
\centering
\subfloat[][Trajectories in the $(x,y)$-plane.]
{\includegraphics[width=6cm,keepaspectratio]{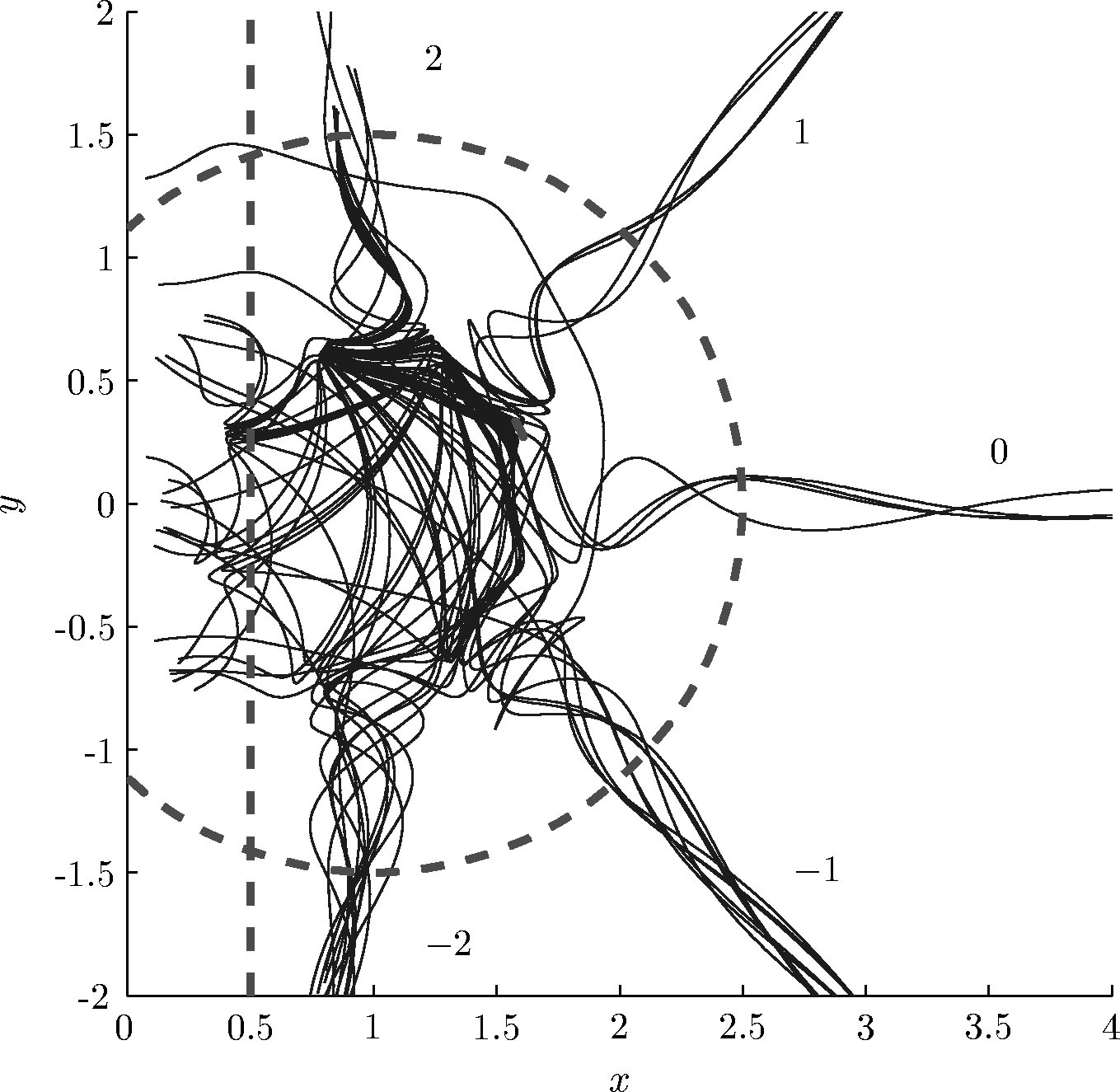}}\qquad
\subfloat[][The function $j(\varphi(i))$.]
{\includegraphics[width=6cm,keepaspectratio]{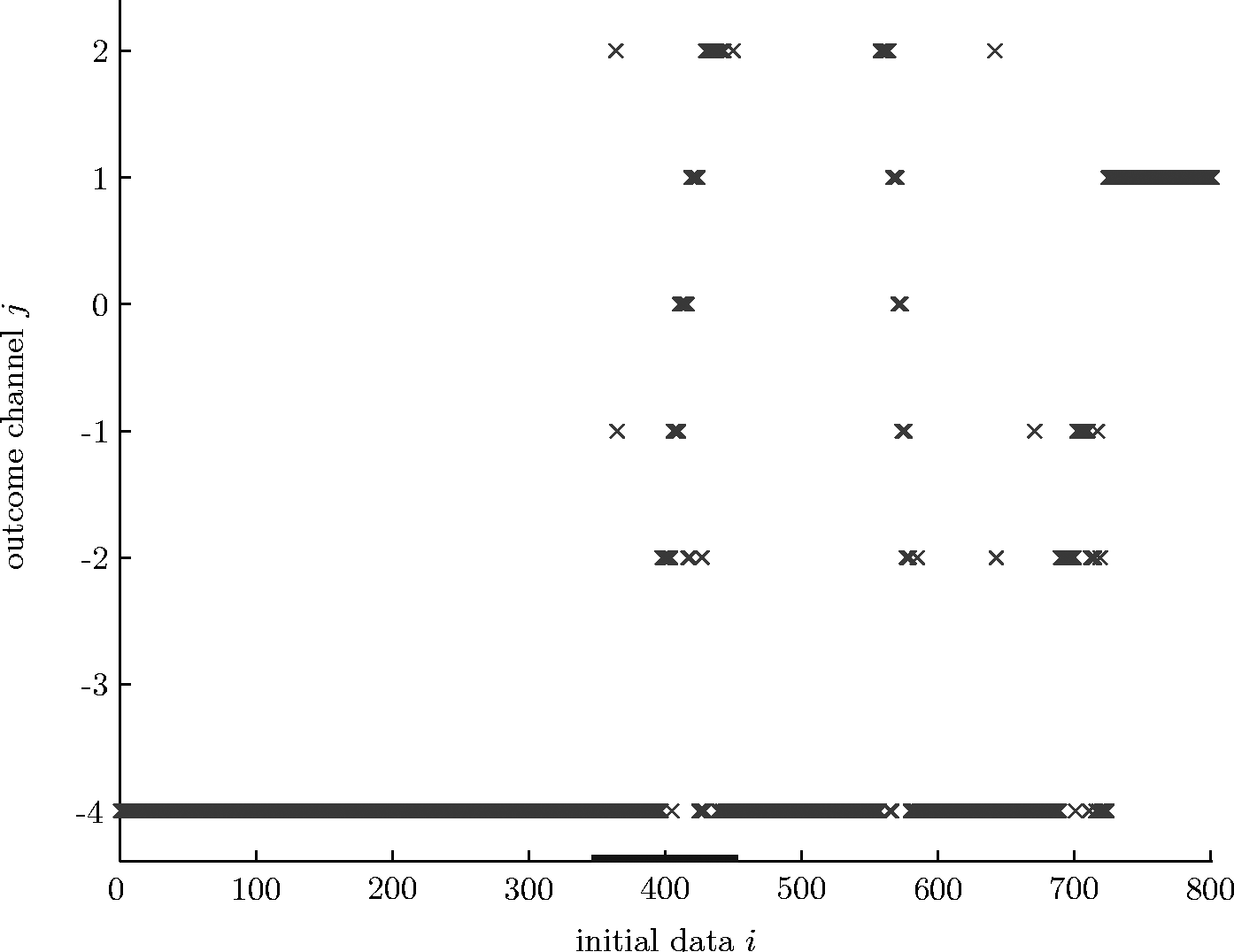}}\\
%\subfloat[][Zoom in the $(x,y)$-plane.]
%{\includegraphics[width=12cm,keepaspectratio]{rc81s512cl1z-b.png}}
\caption{Trajectories of test particles which start from the part of the circle of radius ${r_0=\frac{2}{3}}$ centered around $(x_s,y_s)=(1,0)$. The angular variable $\varphi$ changes discretely from $\frac{2\pi}{15}$ to~$\frac{\pi}{6}$: there are 800 particles labeled by ${i=1,2,\ldots,800}$ whose initial positions are given by ${\varphi(i)=\frac{2\pi}{15}+\frac{\pi}{24000}i}$. The bold line between ${i=350}$ and ${i=450}$ on the horizontal axis of part~(b) denotes the region which will be studied in detail in the following figure~\ref{f:rc81l2}.}
\label{f:rc81l1}
\end{figure}

A natural parametrization of the unit circle of \emph{initial positions} 
(\ref{inipos}), (\ref{inipos2}) for the geodesics (some of which are 
visualised in figure~\ref{f:r8-1}) is provided by the polar coordinate $\varphi$. 
Obviously, for large values of $\lambda$ there are (at most)~$n$  outcome channels throughout which 
the unbounded geodesics may approach the curvature singularity at infinite values of $r$,
where $n$ is the power of the metric function (\ref{GObecna}). 
These \emph{distinct channels} represent possible outcomes of our system, and we assign them the symbol
$j$. In the illustrative case ${n=8}$, it takes one of the corresponding discrete values, $j\in\{-2, -1, 0, 1, 2\}$. In particular, ${j=0}$ denotes those geodesics which approach infinity through the channel centered around the axis ${y=0}$, ${j=1}$ corresponds to ``upper right'' channel, 
${j=2}$ corresponds to ``upper left'' channel, ${j=-1}$ and ${j=-2}$ are their ``lower'' counterparts,  see part (a) of figure~\ref{f:rc81l1}. 

In addition to these five labeled outcome channels, there are three undistinguishable  channels oriented towards the left. These are influenced by the envelope singularity localized at ${x=0}$ so strongly that we must ignore them as outcome channels: we assign them a common superfluous label ``$-4$''. To  be precise, a~geodesic is assigned the outcome value ${j=-4}$, once the value of its coordinate $x$ drops below 0.5 --- this (somewhat artificial but given) boundary is denoted by a vertical dashed line in  part (a) of figure~\ref{f:rc81l1}. The dashed circle of radius 1.5 indicates the value of $r$ at which the labels $j=-2,-1,0,1,2$ are  assigned to the remaining geodesics, i.e. where the outcome channels are discriminated. Note that the function $j(\varphi)$ is the analogue of the scattering function which is usually introduced in the classical chaotic 
scattering problems \cite{Ott} with unbounded trajectories. 

\begin{figure}[ht]     % -- level 2 -------
\centering
\subfloat[][Trajectories in the $(x,y)$-plane.]
{\includegraphics[width=6cm,keepaspectratio]{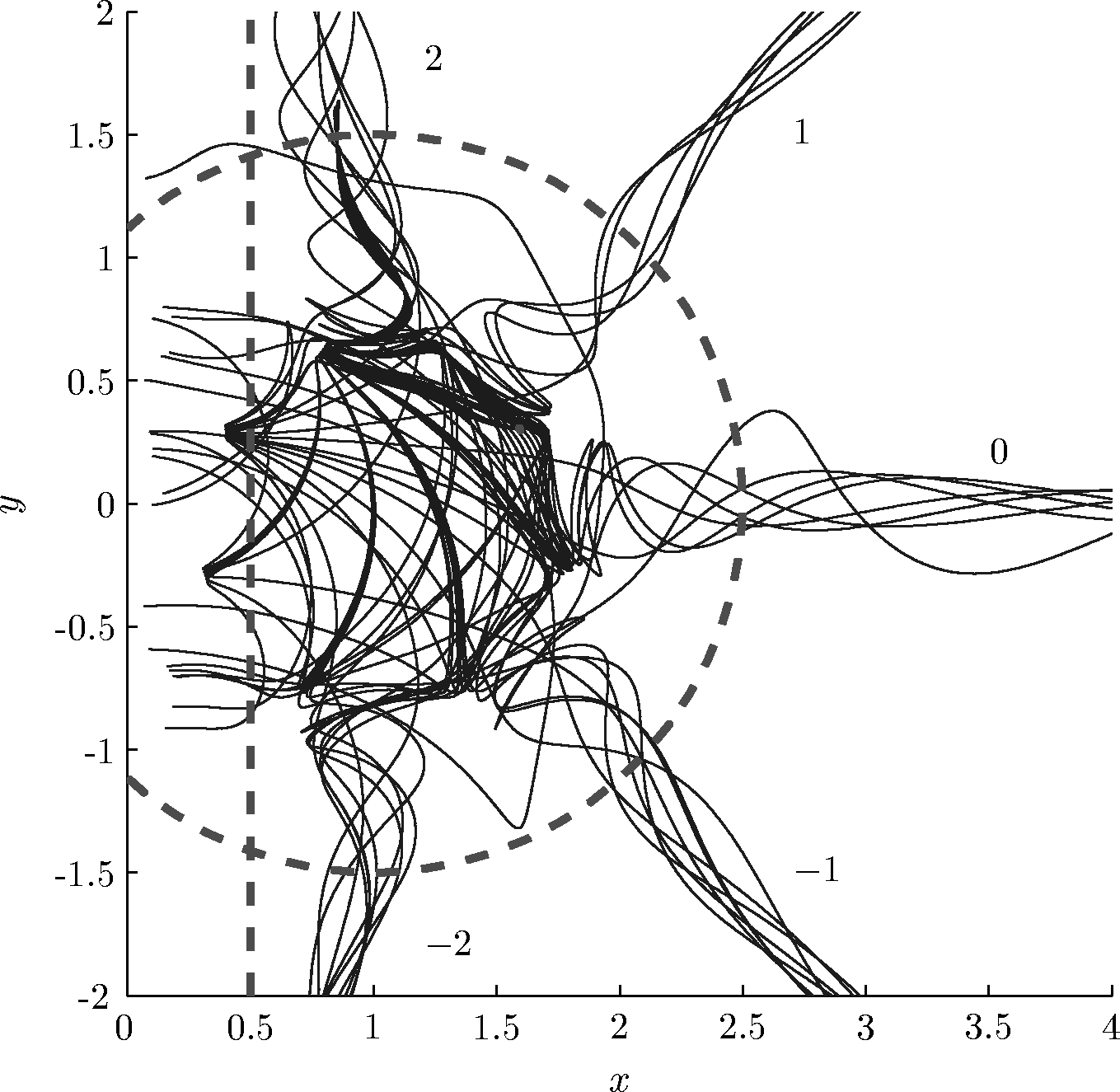}}\qquad
\subfloat[][The function $j(\varphi(i))$.]
{\includegraphics[width=6cm,keepaspectratio]{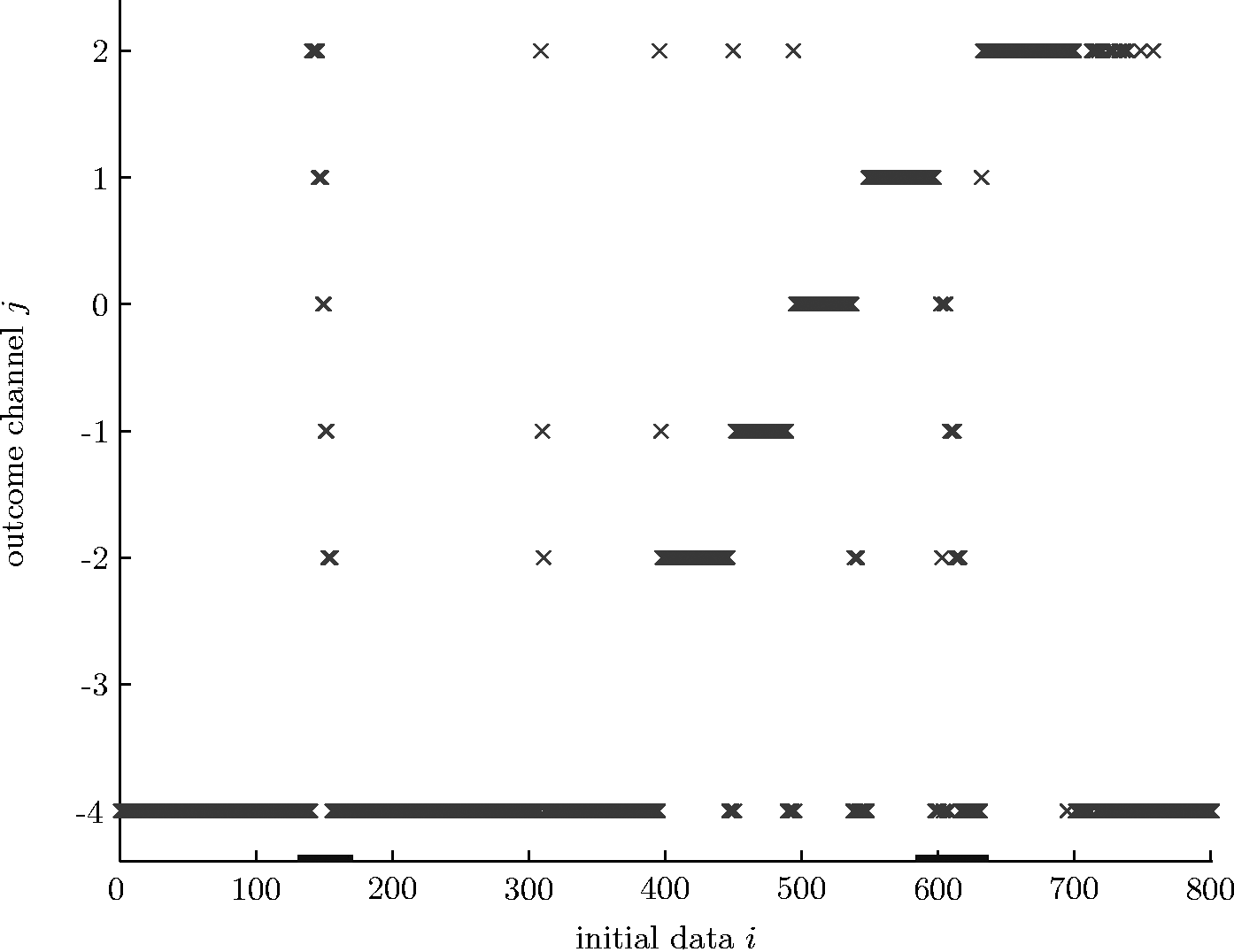}}\\
%\subfloat[][Zoom in the $(x,y)$-plane.]
%{\includegraphics[width=12cm,keepaspectratio]{rc81s512cl2z-b.png}}
\caption{Refinement of the previous figure~\ref{f:rc81l1}.  The angular variable $\varphi$ changes from $\frac{443\pi}{3000}$ to $\frac{1369\pi}{9000}$ according to ${\varphi(i)=\frac{443\pi}{3000}+\frac{\pi}{180000}i}$, where ${i=1,2,\ldots,800}$. The bold lines on the horizontal axis of part (b) depict the two regions studied in detail in  figures~\ref{f:rc81l3a} and ~\ref{f:rc81l3b} (index $i$ running from 131 to 171, and from 584 to 637, respectively).}
\label{f:rc81l2}
\end{figure}

\begin{figure}     % -- level 3a --------
\centering
\subfloat[][Trajectories in the $(x,y)$-plane.]
{\includegraphics[width=6cm,keepaspectratio]{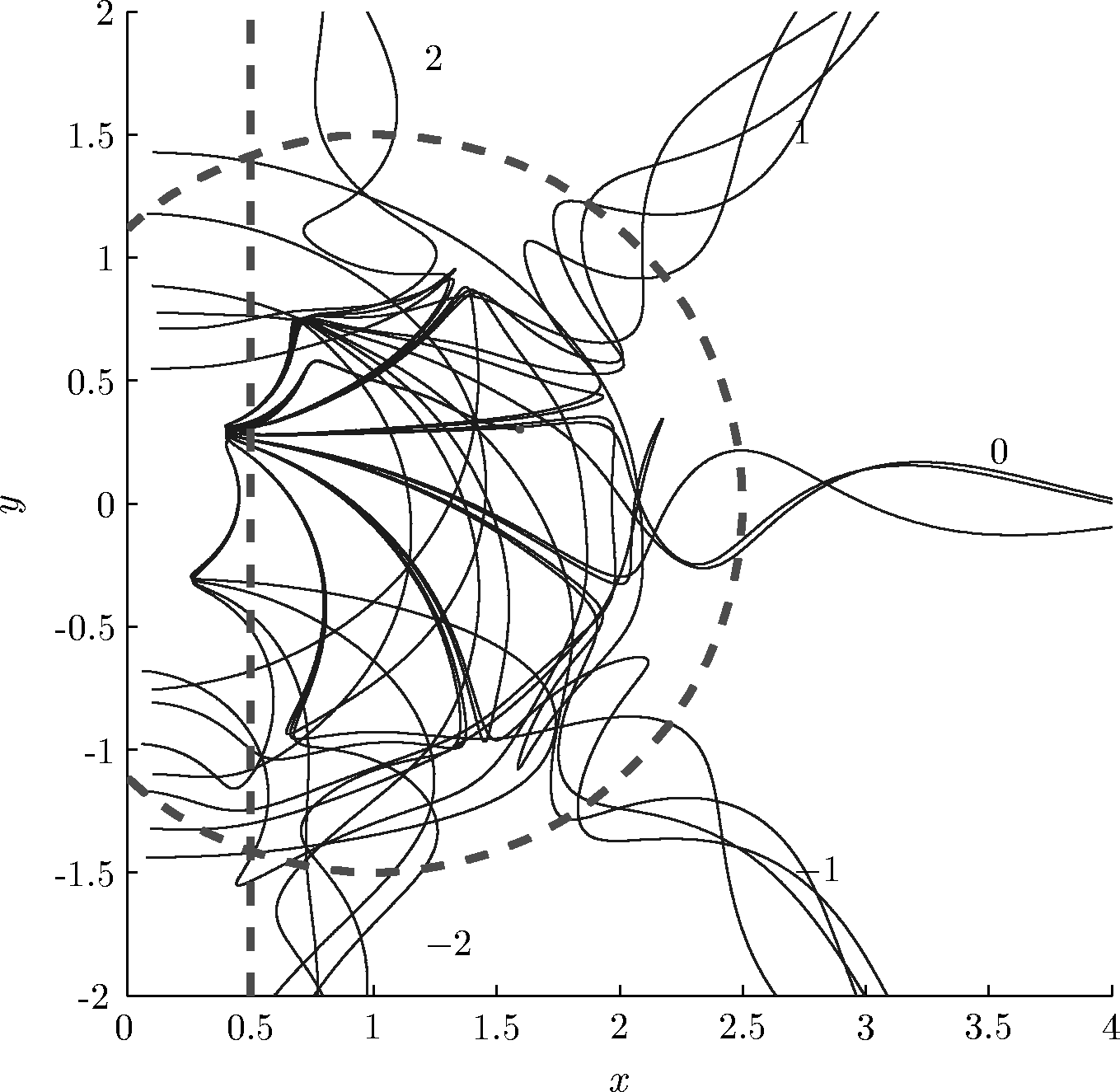}}\qquad
\subfloat[][The function $j(\varphi(i))$.]
{\includegraphics[width=6cm,keepaspectratio]{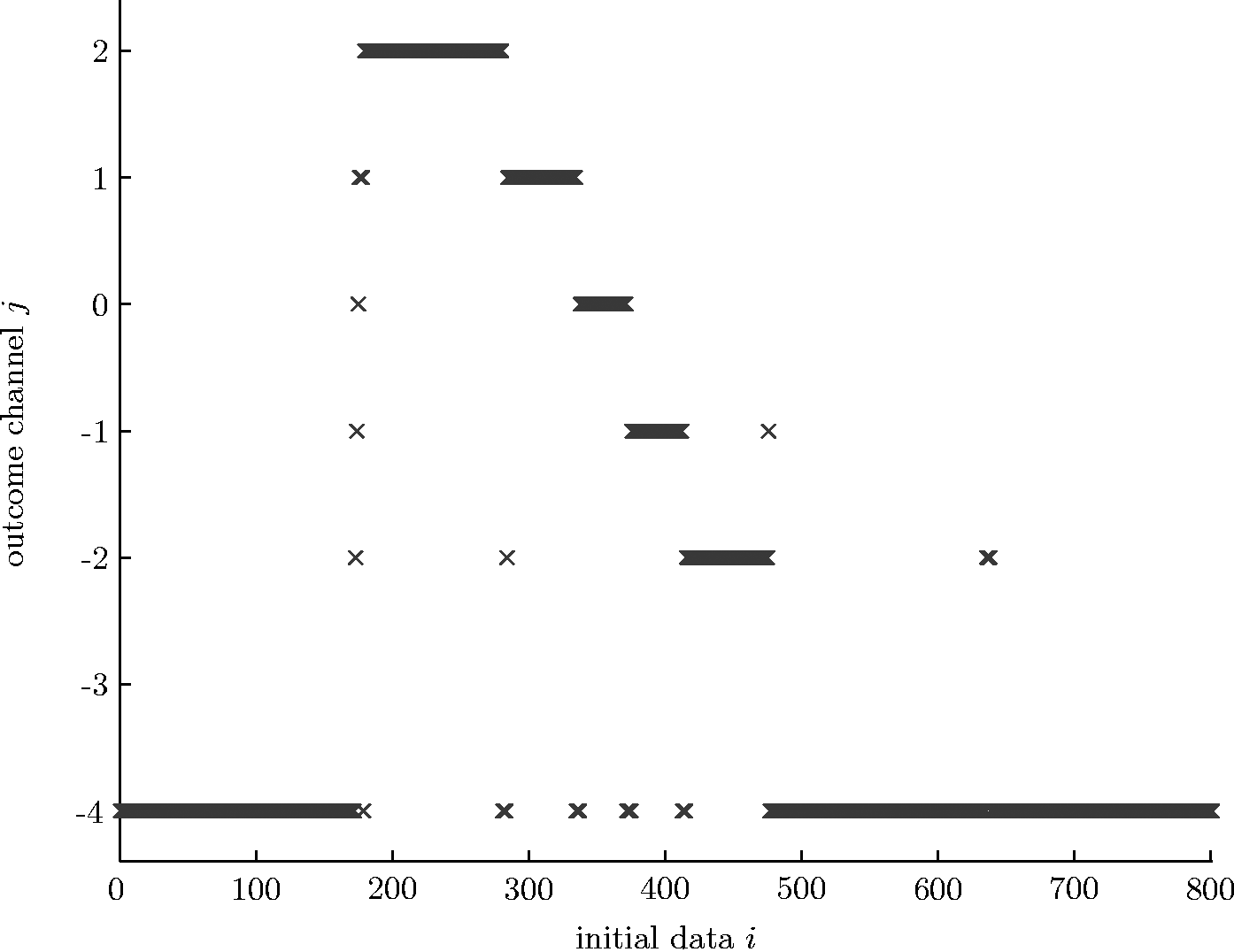}}\\
%\subfloat[][Zoom in the $(x,y)$-plane.]
%{\includegraphics[width=12cm,keepaspectratio]{rc81s512cl3Az-b.png}}
\caption{Refinement of the previous figure~\ref{f:rc81l2}.  The angular variable $\varphi$ changes from $\frac{20033\pi}{135000}$ to $\frac{20063\pi}{135000}$. There are 800 particles numbered by $i$ which start at ${\varphi(i)=\frac{20033\pi}{135000}+\frac{\pi}{3600000}i}$.}
\label{f:rc81l3a}
\end{figure}

\begin{figure}     % -- level 3b -------
\centering
\subfloat[][Trajectories in the $(x,y)$-plane.]
{\includegraphics[width=6cm,keepaspectratio]{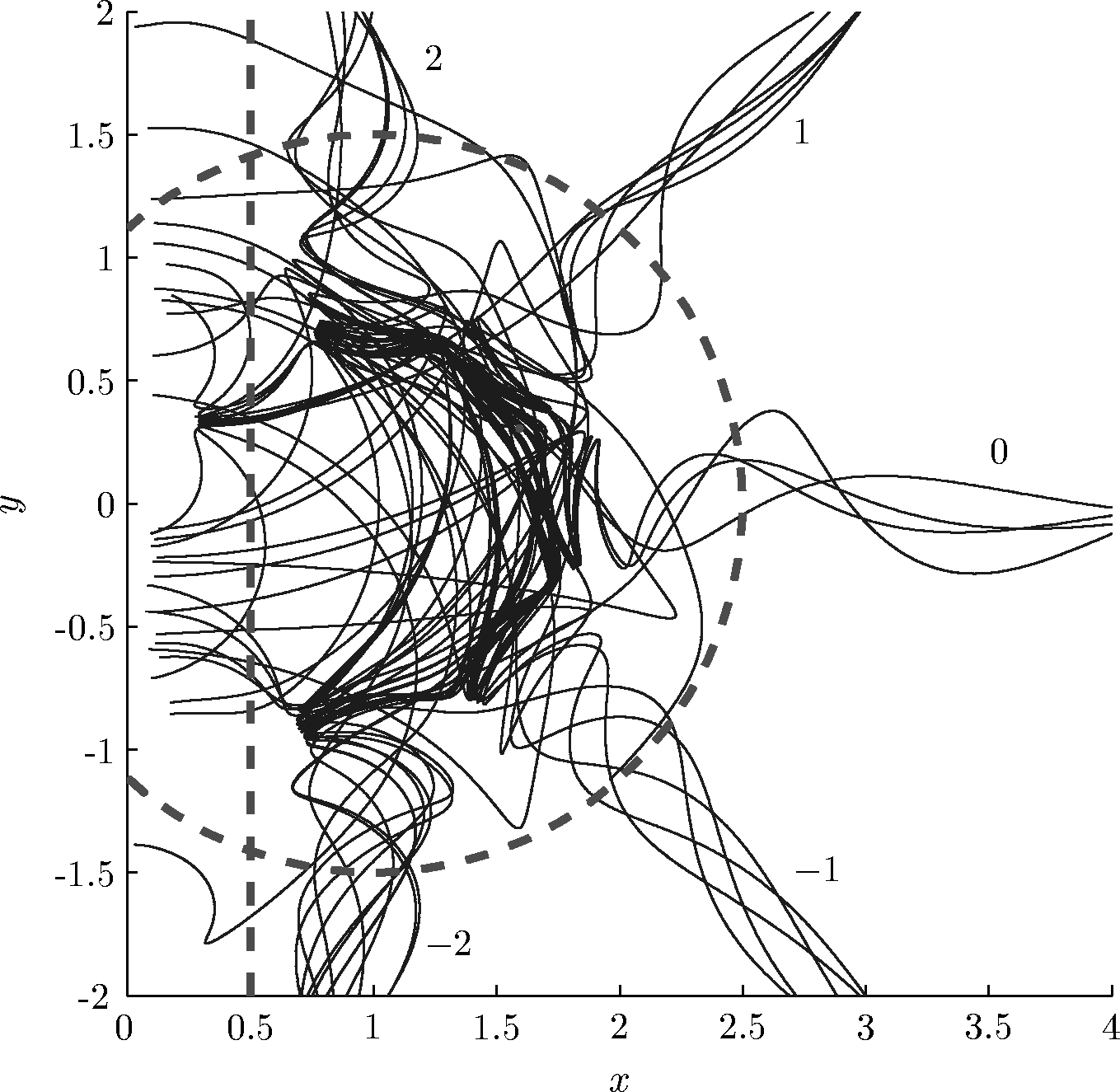}}\qquad
\subfloat[][The function $j(\varphi(i))$.]
{\includegraphics[width=6cm,keepaspectratio]{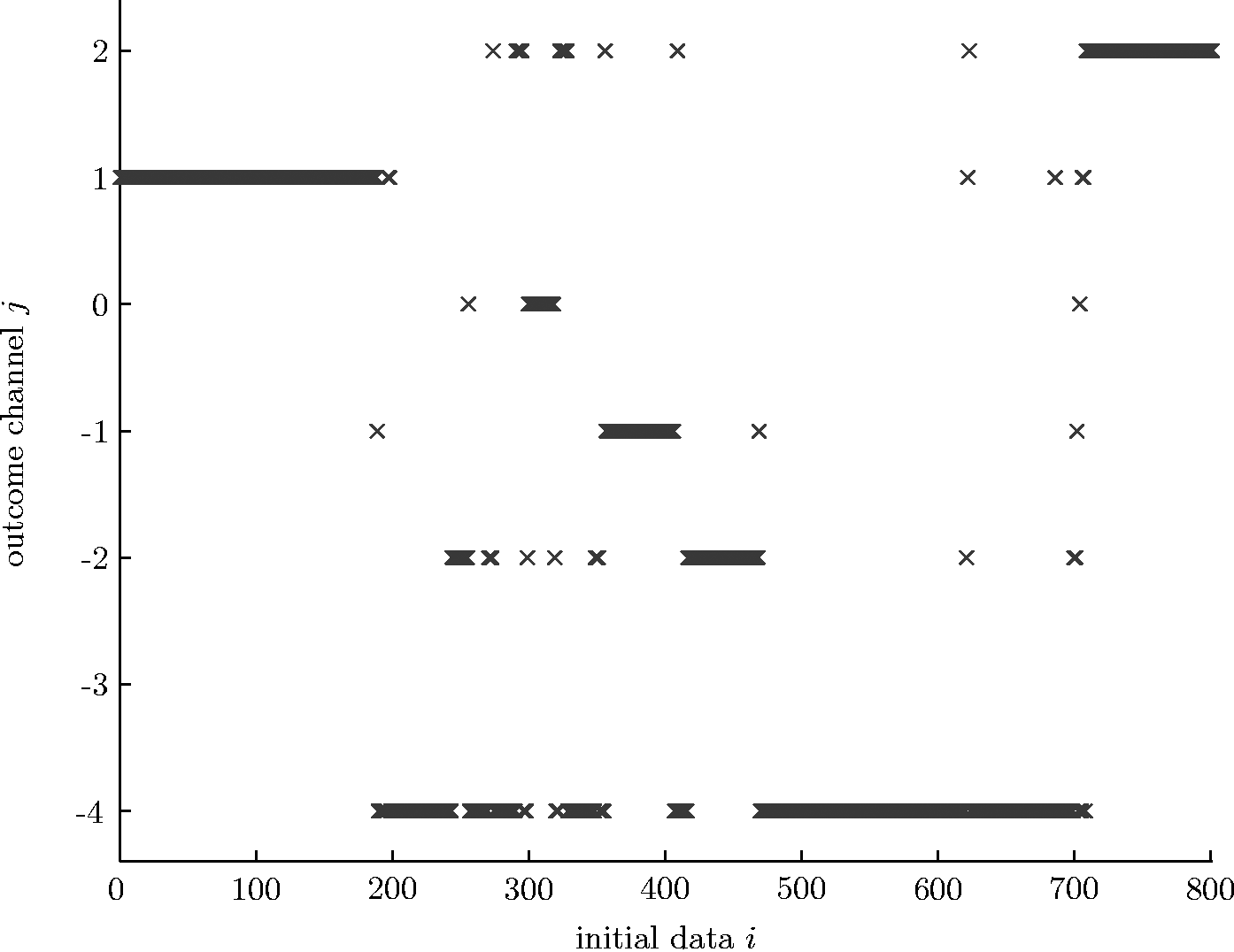}}\\
%\subfloat[][Zoom in the $(x,y)$-plane.]
%{\includegraphics[width=12cm,keepaspectratio]{rc81s512cl3Bz-b.png}}
\caption{Different refinement of the figure~\ref{f:rc81l2}.  The angular variable $\varphi$ changes from $\frac{6791\pi}{45000}$ to $\frac{20413\pi}{135000}$, there are 800 particles numbered by $i$ which start at ${\varphi(i)=\frac{6791\pi}{45000}+\frac{\pi}{2700000}i}$.}
\label{f:rc81l3b}
\end{figure}

 We observe from the right part (b) of figure~\ref{f:rc81l1} that in certain regions the behaviour of the function $j(\varphi)$ \emph{depends very sensitively on} $\varphi$. As a typical example we present $j(\varphi(i))$ calculated numerically for 800 geodesics labeled by $i$ whose positions on the initial circle ${r_0=\frac{2}{3}}$ (shifted by ${x_s=1}$ in the spacetime with $\lambda=512$ and $n=8$) were parametrized by ${\varphi\in(\frac{2\pi}{15},\frac{\pi}{6})=(0.4189,0.5236)}$ such that ${\varphi(i)=\frac{2\pi}{15}+\frac{\pi}{24000}i}$, where ${i=1,2,\ldots, 800}$. We display some of the corresponding trajectories in the left part (a) of figure~\ref{f:rc81l1}.
In fact, the basin boundaries between the outcomes --- indicated by the discontinuities in the function $j(\varphi)$ --- appear to be \emph{fractal}.
This is confirmed on the enlarged detail of the part~(b) of figure~\ref{f:rc81l1}, and on the
enlarged detail of this detail in figures~\ref{f:rc81l2}, \ref{f:rc81l3a} and~\ref{f:rc81l3b}, respectively. 


In figure~\ref{f:rc81l2} we show such zooming in the interval ${\varphi\in(\frac{443\pi}{3000},\frac{1369\pi}{9000})=(0.4639,0.4779)}$ where we plot 800 geodesics initially located at ${\varphi(i)=\frac{443\pi}{3000}+\frac{\pi}{180000}i}$. In figure~\ref{f:rc81l3a} there is a deeper magnification covering the smaller interval ${\varphi\in(\frac{20033\pi}{135000},\frac{20063\pi}{135000})=(0.4662,0.4669)}$ by 800 geodesics starting on   ${\varphi(i)=\frac{20033\pi}{135000}+\frac{\pi}{3600000}i}$, and similarly in figure~\ref{f:rc81l3b} we present results for another 800 geodesics which start in the interval ${\varphi\in(\frac{6791\pi}{45000},\frac{20413\pi}{135000})=(0.4741,0.4750)}$ with initial positions ${\varphi(i)=\frac{6791\pi}{45000}+\frac{\pi}{2700000}i}$. From these typical pictures we conclude that the basin boundaries separating the initial data leading to different discrete output channels $j$ are highly complicated and, in fact, appears to have a fractal structure. The dependence of the outcome on the initial condition is extremely sensitive. 

It thus seems to be a plausible conjecture that the geodesic motion in the Kundt gravitational wave spacetimes studied is genuinely chaotic, 
as in the previously investigated case of other spacetimes  \cite{Conto1,DFC,Yur,CG}, or \cite{PVcha,PoVe2,PoVe3,VePo}. However, to rigorously prove this statement would require a demonstration that the basin  boundaries are fractal to an \emph{arbitrary high level}, i.e. for differences in the initial data which approach zero. Of course, this cannot be achieved  solely by numerical integration of geodesics due to the inevitable  numerical errors. In the case of chaotic motion in non-homogeneous {\it pp}-waves [18-21] the numerical investigations were complemented and supported  by analytic methods: in the transverse section spanned by the ${(x,y)}$ spatial coordinates the evolution equations are equivalent to a Hamiltonian system with  H\'{e}non--Heiles-type or more general $n$-saddle polynomial potentials which are ``textbook'' examples of chaotic systems \cite{Rod, RPC, Churchil}. Unfortunately, in the present case of motion in the Kundt spacetimes, no such analytic works are available. The reason is that the system of equations (2)--(5) is much more complex. In particular, the motion in the transverse ${(x,y)}$ section is \emph{not} decoupled from the $u$ and $v$ coordinates.

\begin{figure}     % -- all details -----
\centering
%\subfloat[][Trajectories in the $(x,y)$-plane.]
%{\includegraphics[width=6cm,keepaspectratio]{rc81s512cl1.png}}\qquad
%\subfloat[][The function $j(\varphi(i))$.]
%{\includegraphics[width=6cm,keepaspectratio]{rc81s512cl1c.png}}\\
\subfloat[][Detail of the figure~\ref{f:rc81l1}.]
{\includegraphics[width=12cm,keepaspectratio]{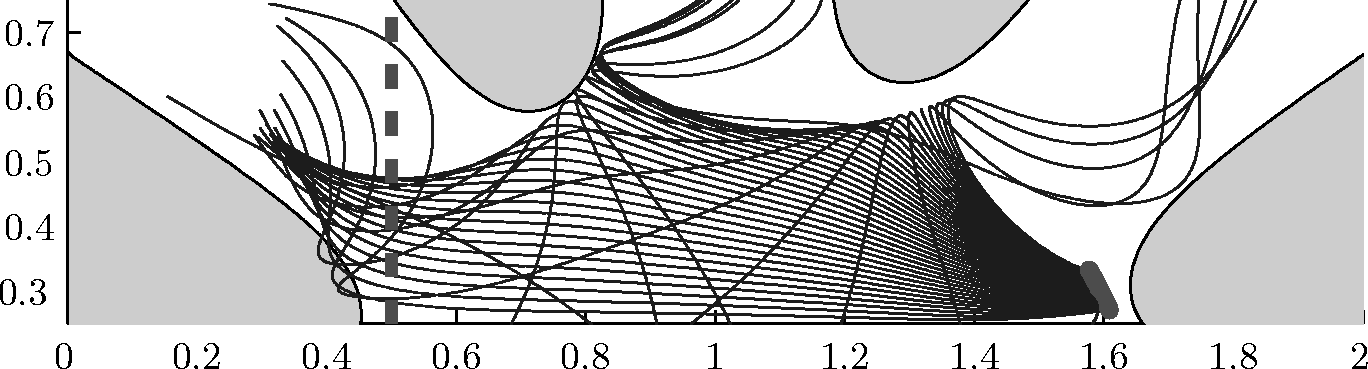}}\\
\subfloat[][Detail of the figure~\ref{f:rc81l2}.]
{\includegraphics[width=12cm,keepaspectratio]{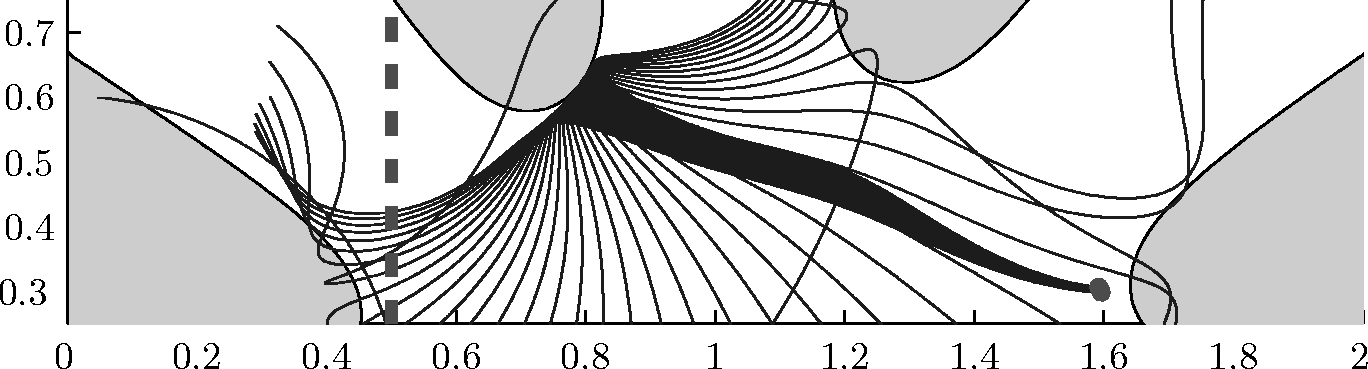}}\\
\subfloat[][Detail of the figure~\ref{f:rc81l3a}.]
{\includegraphics[width=12cm,keepaspectratio]{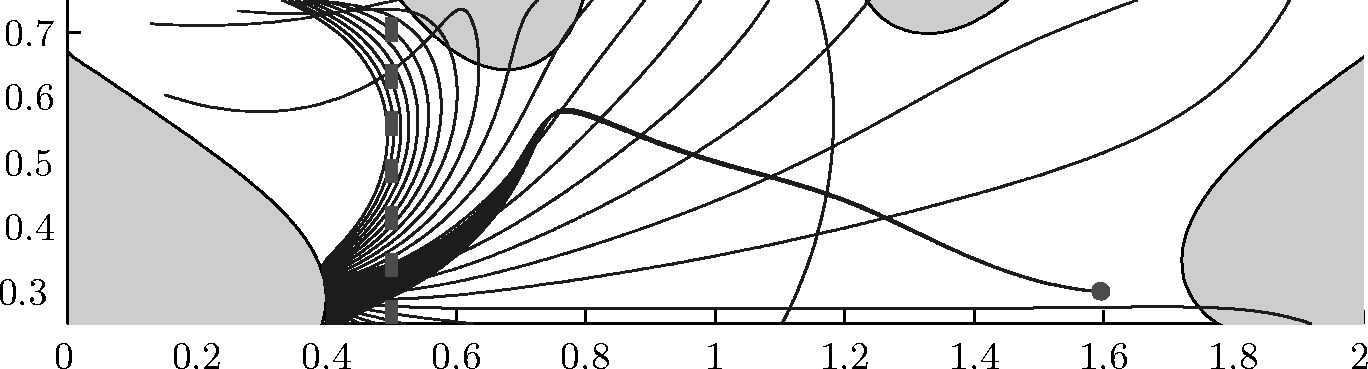}}\\
\subfloat[][Detail of the figure~\ref{f:rc81l3b}.]
{\includegraphics[width=12cm,keepaspectratio]{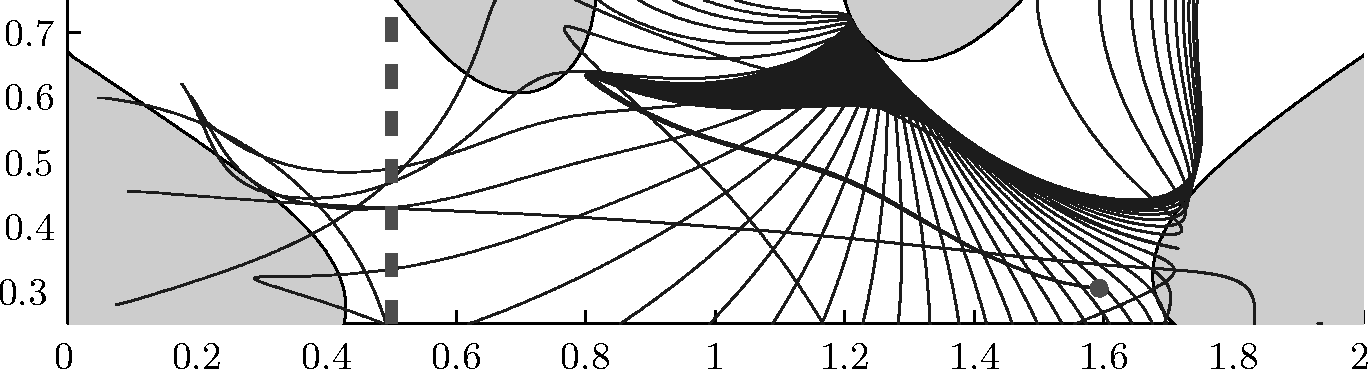}}
\caption{Zoom in the $(x,y)$-plane  which shows details of parts (a) of figures~\ref{f:rc81l1},~\ref{f:rc81l2},~\ref{f:rc81l3a} and~\ref{f:rc81l3b}. The geodesic test particles effectively bounce on concave boundaries which results in a rapid dispersion of the trajectories and their fractal separation. The higher the level of the fractal, the greater the number of bounces that a particle undergoes before it escapes to infinity.}
\label{f:details}
\end{figure}

We can thus present only the following heuristic argument for the presence of a genuine chaos. From our numerical simulations we observe that in the ${(x,y)}$ section the geodesics are asymptotically restricted to the regions in which the Killing vector $\partial_u$ is spacelike, see figure~\ref{f:kill-re}. Moreover, the geodesics ``bounce'' on the corresponding boundaries, as indicated on figure~\ref{f:details} which shows the details of figures~\ref{f:rc81l1},~\ref{f:rc81l2},~\ref{f:rc81l3a} and~\ref{f:rc81l3b}. Let us note that the boundaries also depend on the value of  the coordinate $v$ and they therefore evolve with time. In the central region these geometric boundaries effectively form walls with (time dependent) concave shapes, and it is also natural to expect that the geodesic particles may bounce arbitrarily many times  between them before they  are scattered to one of the outcome channels labeled by $j$. The curved wall segments have a ``dispersing effect'' on the incident trajectories, and this leads to a rapid separation  of nearby geodesics after a few bounces. We thus expect that the basin boundaries separating the outcomes are fractal to \emph{any level} (the level corresponding basically to the number of bounces performed between the walls in the central region), as in the familiar chaotic billiard systems \cite{Ott}.

\newpage
\section{Conclusions}

Some time ago, the first explicit demonstration of chaos in exact radiative spacetimes
was given  for a family of non-homogeneous {\it pp}-wave solutions 
which represent the simplest models of exact gravitational waves in general relativity
\cite{PVcha,PoVe2,PoVe3,VePo}. In the present paper we have extended these results. 
Using the invariant fractal method we have numerically demonstrated a complicated chaotic character of 
geodesic motion in the class of Kundt spacetimes (\ref{metric}) with polynomial metric function (\ref{GObecna}). 
These vacuum spacetimes are of type~N, and therefore they also represent exact gravitational waves.

Our work shows that geodesic motion in spacetimes even as simple  as {\it pp}-waves and the Kundt waves can be complex for specific but natural choices of the corresponding metric functions. This may possibly stimulate future investigation of chaotic behaviour of geodesics in other exact radiative spacetimes, such as non-expanding Kundt waves with a cosmological constant.

\section*{Acknowledgments}

We acknowledge the support of the grant GA\v{C}R~202/06/0041 and the Czech Centre for Theoretical Astrophysics LC06014. We wish to thank Jerry Griffiths for reading the manuscript.


\vspace{2mm}


\begin{thebibliography}{99}
\setlength{\itemsep}{-1mm}

\bibitem{Conto1} Contopoulos~G 1990
    {\it Proc.~R.~Soc.~Lond.} A {\bf 431} 183;
    1991 {\bf 435} 551

\bibitem{BoCa} Bombelli~L and Calzetta~E 1992
    {\it Class.~Quantum~Grav.} {\bf 9} 2573

\bibitem{DFC} Dettmann~C~P, Frankel~N~E and Cornish~N~J 1994
    {\it Phys.~Rev.}~D {\bf 50} R618

\bibitem{Yur} Yurtsever~U 1995
    {\it Phys.~Rev.}~D {\bf 52} 3176

\bibitem{CG}  Cornish~N~J and Gibbons~G~W 1997
    {\it Class.~Quantum~Grav.} {\bf 14} 1865

\bibitem{LeVi} Letelier~P~S and Vieira~W~M 1997
    {\it Class.~Quantum~Grav.} {\bf 14} 1249

\bibitem{SuMa} Suzuki~S and Maeda~K 1997
    {\it Phys.~Rev.}~D {\bf 55} 4848
    
\bibitem{Le1} Levin~J 2000
    {\it Phys.~Rev.~Lett.} {\bf 84} 3515
    
\bibitem{Le2} Levin~J 2003
    {\it Phys.~Rev.}~D {\bf 67} 044013
    
\bibitem{CL}  Cornish~N~J and Levin~J  2003
    {\it Class.~Quantum~Grav.} {\bf 20} 1649   

\bibitem{Hartl1} Hartl~M~D 2003
    {\it Phys.~Rev.}~D {\bf 67} 024005    

\bibitem{Hartl2} Hartl~M~D 2003
    {\it Phys.~Rev.}~D {\bf 67} 104023    

\bibitem{HaBu} Hartl~M~D and Buonanno~A 2005
    {\it Phys.~Rev.}~D {\bf 71} 024027    

\bibitem{KaVo} Karas~V and Vokrouhlick\'y D 1992
    {\it Gen.~Rel.~Grav.} {\bf 24} 729

\bibitem{SSM} Sota~Y, Suzuki~S and Maeda~K 1996
    {\it Class.~Quantum~Grav.} {\bf 13} 1241

\bibitem{ViLe} Vieira~W~M and Letelier~P~S 1996
    {\it Phys.~Rev.~Lett.} {\bf 76} 1409

\bibitem{ViLe2} Vieira~W~M and Letelier~P~S 1996
    {\it Class.~Quantum~Grav.} {\bf 13} 3115

\bibitem{PVcha}  Podolsk\'y~J and Vesel\'y~K 1998
    {\it Phys.~Rev.}~D {\bf 58} 081501
    
\bibitem{PoVe2} Podolsk\'y~J and Vesel\'y~K 1998
    {\it Class.~Quantum~Grav.} {\bf 15} 3505

\bibitem{PoVe3} Podolsk\'y~J and Vesel\'y~K 1999
    {\it Class.~Quantum~Grav.} {\bf 15} 3599

\bibitem{VePo} Vesel\'y~K and Podolsk\'y~J  2000
    {\it Phys.~Lett.}~A {\bf 271} 368
  
\bibitem{VaPa}  Varvoglis~H and Papadopoulos~D 1992
    {\it Astron.~Astrophys.}  {\bf 261} 664

\bibitem{CF}   Cornish~N~J and Frankel~N~E 1997
    {\it Phys.~Rev.}~D {\bf 56} 1903

\bibitem{CMR} Chicone~C, Mashoon~B and Retzloff~D~G 1997
    {\it Class.~Quantum~Grav.} {\bf 14} 699

\bibitem{Koku} Kokubun~F 1998
    {\it Phys.~Rev.}~D {\bf 57} 2610

\bibitem{KSMH} Stephani~H, Kramer~D, MacCallum~M, Hoenselaers~C and Herlt~E
   2003  {\it Exact solutions of Einstein's field equations, 2nd edition}
   (Cambridge: Cambridge University Press)
    
\bibitem{Kundt61} Kundt W 1961
    {\it Z. Phys.\/} {\bf 163} 77

\bibitem{Kundt62} Kundt W 1962
    {\it Proc. R. Soc. {\rm A}\/} {\bf 270} 328

\bibitem{EK} Ehlers J and Kundt W 1962
   {\it Gravitation: An Introduction to Current Research\/}, ed L~Witten (New York: Wiley) p 49
    
\bibitem{PodBel04}  Podolsk\'y~J and Bel\'a\v n~M 2004
    {\it Class.~Quantum~Grav.} {\bf 21} 2811

\bibitem{BicakPravda}
Bi\v c\' ak J and Pravda   V 1998 {\em Class. Quantum Grav.\/} {\bf 15} 1539

\bibitem{Pravdovi}
Pravda V, Pravdov\'a A, Coley A and Milson R 2002 {\em Class. Quantum Grav.\/}
  {\bf 19} 6213

\bibitem{Coley}
Coley A~A 2002 {\em Phys. Rev. Lett.\/}  {\bf 89} 281601

\bibitem{BicPod99II}
Bi\v{c}\'ak J and Podolsk\'y J 1999 {\em J. Math. Phys.\/} {\bf 40} 4506

\bibitem{GriPod98}  Griffiths~J~B and Podolsk\'y~J 1998
    {\it Class.~Quantum~Grav.} {\bf 15} 3863

\bibitem{Ott} Ott E 1993
   {\it Chaos in Dynamical Systems} (Cambridge: Cambridge University Press)

\bibitem{Rod} Rod D L 1973 {\em J. Differ. Equ.\/} {\bf 14} 129

\bibitem{RPC} Rod D L, Pecelli G and Churchil R C 1977 {\em J. Differ. Equ.\/} {\bf 24} 329

\bibitem{Churchil} Churchil R C and Rod D L 1980 {\em J. Differ. Equ.\/} {\bf 37} 23


\end{thebibliography}
\end{document}